\documentclass[runningheads]{llncs}
\usepackage{graphicx}
\usepackage{caption}
\usepackage{subcaption}
\PassOptionsToPackage{hyphens}{url}\usepackage{hyperref}

\begin{document}
\title{Google Topics as a way out of the cookie dilemma?}
%
%
\author{Marius Köppel\inst{1}, Jan-Philipp Muttach (n\'e Stroscher)\inst{2} \and Gerrit Hornung\inst{2}}
\authorrunning{M. Köppel and J. P. Muttach (n\'e Stroscher) and G. Hornung}
%
\institute{Johannes Gutenberg Universität, Saarstraße 21, 55122 Mainz, Germany \email{mkoeppel\{@\}uni-mainz.de} \and
Universität Kassel, Henschelstraße 4, 34127 Kassel, Germany
\email{jan-philipp.muttach\{@\}uni-kassel.de}\\
\email{gerrit.hornung\{@\}uni-kassel.de}
}
\maketitle              
\setcounter{footnote}{0}

\begin{abstract}
The paper discusses the legal requirements and implications of the processing of information and personal data for advertising purposes, particularly in the light of the "Planet49" decision of the European Court of Justice (ECJ) and the "Cookie Consent II" decision by the German Federal Court (Bundesgerichtshof, BGH).
It emphasises that obtaining explicit consent of individuals is necessary for setting cookies.
The introduction of the German Telecommunication Telemedia Data Protection Act (Telekommunikation-Telemedien-Datenschutzgesetz, TTDSG) has replaced the relevant section of the German Telemedia Act (Telemediengesetz, TMG) and transpose the concept of informed consent for storing and accessing information on terminal equipment, aligning with Article 5(3) ePrivacy Directive.
To meet these requirements, companies exploring alternatives to obtaining consent are developing technical mechanisms that rely on a legal basis.
Google tested initially "Federated Learning of Cohorts" (FLoC) as part of their "Privacy Sandbox" strategy.
This technology was significantly criticized, Google introduced a new project called "Google Topics", which aims to personalize advertising by categorizing users into interest groups, called topics.
Implementation of this technology began in July 2023.

\end{abstract}
\section{Introduction}

\begin{figure}
    \centering
    \begin{subfigure}[b]{0.49\textwidth}
        \centering
        \includegraphics[width=\textwidth]{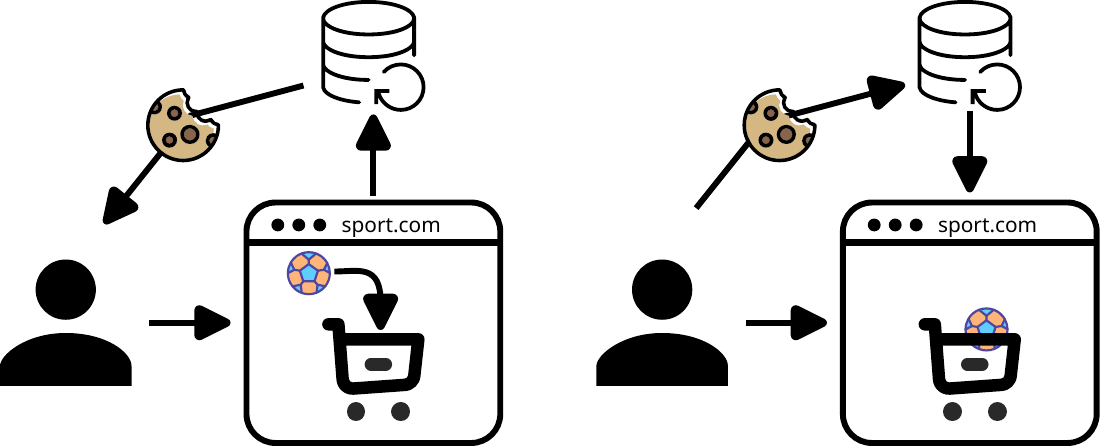}
        \caption{First party cookies.}
        \label{fig:first}
    \end{subfigure}
    \hfill
    \begin{subfigure}[b]{0.49\textwidth}
        \centering
        \includegraphics[width=\textwidth]{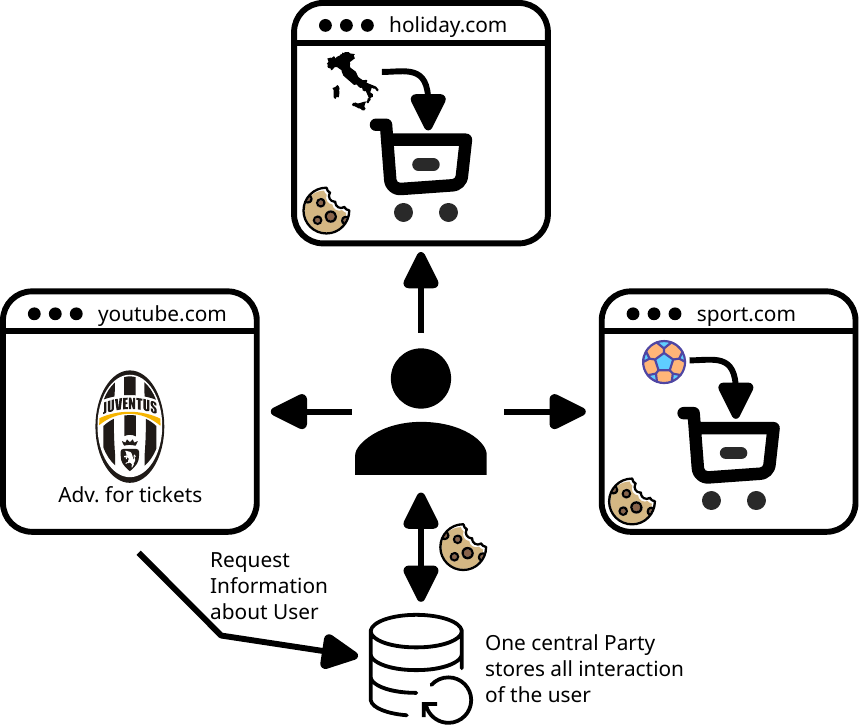}
        \caption{Third party cookies.}
        \label{fig:third}
    \end{subfigure}
    \hfill
    \caption{Figure~\ref{fig:first} shows the usage of first party cookies to handle basic website functionalities, while Figure~\ref{fig:third} shows the usage of third party cookies to establish user profiles for advertising purposes. The logo from Juventus Turin was taken from~\cite{turin}.}
        \label{fig:cookies}
\end{figure}
Since the "Planet49" ~\cite{Planet49} and "Cookie Consent II" decisions ~\cite{Planet492}, it has been clear that the setting of so-called third party cookies (see Figure~\ref{fig:cookies} for the basic concepts of first and third party cookies) for advertising purposes requires the explicit consent of the persons concerned.
With the entry into force of the TTDSG on 1 December 2021, Section 15 (3) TMG was replaced by Section 25 TTDSG~\footnote{Section 25 (1) TTDSG: "The storage of information in the terminal equipment of the end-user or the access to information already stored in the terminal equipment shall only be allowed if the end-user has given consent on the basis of clear and comprehensive information.
The information to the end-user and the consent shall be provided in accordance with Regulation (EU) 679/2016."}, which generally requires informed consent for the storage of and access to information on terminal equipment and thus transposes Article 5 (3) ePrivacy Directive (2009/136/EC) into German law for the first time.
If companies want to continue to display personalized advertising through real-time bidding~\cite{bidding1,bidding2,bidding3}, they must obtain informed consent~\cite{bidding1,dark-bannern}.
\\ \\
In order to be independent of this requirement, work is being done on technical mechanisms that are equally effective, but rely on a legal basis, thus eliminating the need for consent~\cite{heise-cookie-end}.
One such technology is Federated Learning of Cohorts (FLoC), which has been tested by Google since 2021.
It was part of the "Privacy Sandbox"~\cite{google-sandbox} strategy to improve user privacy, but has only been tested outside the scope of the General Data Protection Regulation (GDPR) and the ePrivacy Directive~\cite{google-blog-privacy}.
Following significant criticism, Google announced that it would not pursue FLoC further.
At the same time, it introduced a new project, "Google Topics" was introduced, which also aims to personalize advertising by assigning users to interest groups (topics) and has similarities to FLoC~\cite{heise-cookie-nachfolge,topics-api}.
Google announced in May 2023 that the Application Programming Interface (API) for the Privacy Sandbox has been completed.
From the Chrome version to be released in July 2023, websites will be able to use the interfaces built into the browser~\cite{topics-is-ready}.

\section{Functioning of FLoC}
FLoC~\cite{dsgvo-floc} was a technology designed to assign users to cohorts using an algorithm within the Google Chrome browser.
Originally, Google intended to use "Federated Learning".
In this approach, the machine learning models would be trained in a decentralized manner on the user’s device, and the updated models would then be sent to a central server for aggregation.
This is in contrast to the traditional approach where data is first sent to a central server to train the model in a uniform manner.
This similarity serves as the basis for understanding Topics.
The user’s installation of the Chrome browser on their device enabled mapping by evaluating browsing history using a hash function~\footnote{A hash function maps input values to fixed-length output values.
It ensures that similar inputs have different outputs and is currently non-invertible, making it impossible to deduce inputs from outputs.} to generate a "cohort ID" and to associate users with interest groups~\cite{dev-floc}.
\\ \\
FLoC used the SimHash algorithm~\cite{similarity}, where the browsing history was used as the input to calculate a hash value, which was then stored on the user’s computer.
Unlike a general hash function, this algorithm had the property of mapping similar input values to similar hash values, which Google had previously used to detect website duplicates~\cite{duplicates-web}.
This property was to be used in FLoC to allow categorization into cohorts based on similar hash values indicating similar interests.
Website operators and advertising networks were supposed to receive the IDs for different cohorts via an API in Google Chrome.
\\ \\
Anonymity was to be achieved by the inability to re-identify individuals due to the size of the respective cohorts.
However, empirical analysis of user data showed that individuals could be re-identified with over 95\% probability, as certain cohorts allowed inferences to be made about demographics when the cohort sequence was recorded over four weeks and users visited up to 100 websites~\cite{ana-floc}.
The researchers reconstructed the cohorts using a freely available version of the SimHash algorithm, verified by Google.
This finding is based solely on a correlation analysis between cohorts and demographic data.
The probability of re-identification increases when additional information from fingerprinting~\footnote{In fingerprinting, user tracking is not done via third party cookies.
Instead, a profile of users is created based on the hardware, software, browser settings, add-ons, and other characteristics they use.} is included, especially when unique settings such as specific fonts are used~\cite{John.2022,Haberer.2020}.

\section{How Google Topics works and differences to FLoC and cookies}

\begin{figure}
    \centering
    \includegraphics[width=0.7\textwidth]{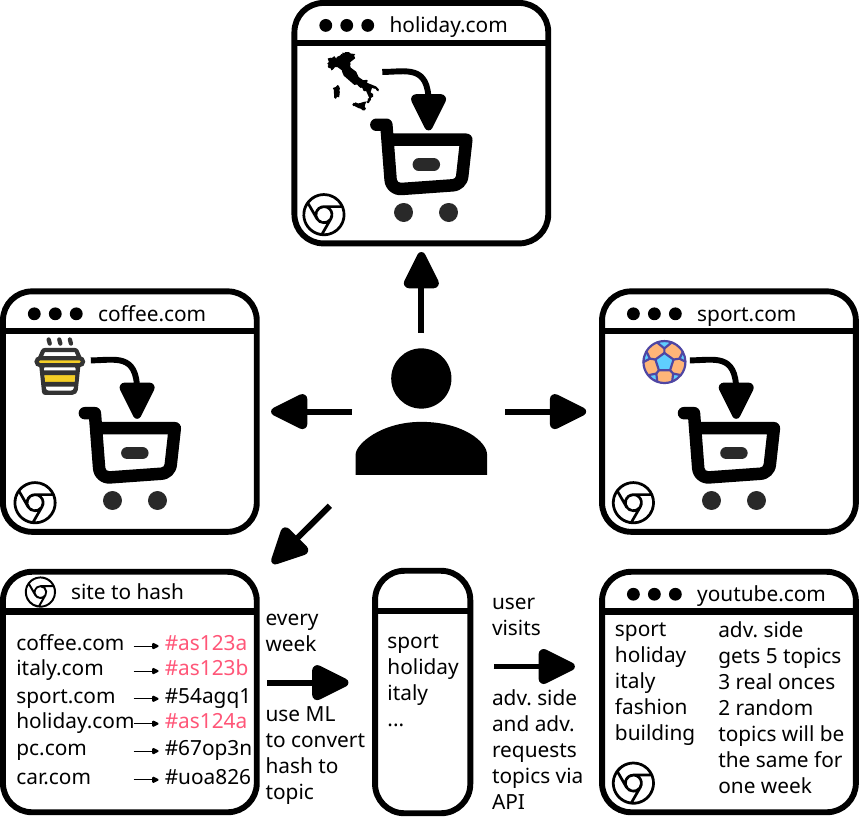}
    \caption{Simplified sketch of the working principle of Google Topics.}
    \label{fig:topics}
\end{figure}
Installation and use of the Google Chrome browser allows categorization by Google Topics~\cite{taxonomyv1,topics-api-guide}.
Google Topics have a predefined set of topics, such as fitness and travel, with further specified subgroups.
Initially, about 350 topics have been defined.
The browser determines topics when a web page is accessed, using a classification model that assigns web pages to topics.
Website owners have the option to provide meta-information that is embedded in the source code of the website (e.g. indicating that it is a football website) to improve the accuracy of topic assignment.
To enable topic identification, a code from the advertising platform (referred to in the documentation as "adtech") must be embedded in each web page visited.
This code queries topics using the \textbf{document.browsingTopics()} function via the Topics API.
In order for advertising platforms to receive topics, the relevant web page must integrate the advertising platform code, which then calls the Topics API.
Once this is done, the execution of the advertising platform code on the website triggers the loading of additional code from the advertising platform.
As there is necessarily communication between the user~\footnote{The term "user" refers to individuals who have installed Google Chrome and use it to browse websites.
It is synonymous with the legally defined term "end user" which refers to the ultimate consumer or user of a product or service.} and the advertising platform during this process, the advertising platform is aware of the user's IP address.
Figure~\ref{fig:topics} shows an outline of the functionality of Topics using the SimHash algorithm.
\\ \\
When a user visits a web page where the advertising platform's code is integrated to request the Topics API (as described above), the page is associated with the user's interest.
This interest is now considered "observed" for that specific ad platform (according to the documentation).
When the same user visits another web page where the Topics API is called by the same advertising platform, the advertising platform can obtain the previously determined topic through the Topics API request.
Similarly, the topic determined based on the visit to that web page is now considered ”observed" for that advertising platform.
The entity making the request to the browser (typically an advertising platform) is called the Caller.
\\ \\
Based on the users’ visits to the website, the top five topics per user are determined each week.
This allocation is updated every week (assuming a sufficient number of site visits) and the topics are only stored for three weeks, resulting in continuous re-allocations.
A random topic is  added to each API query, which the aim of giving the caller a 5\% chance of receiving a topic that does not match the user’s interests.
This is done to make tracking and profiling more difficult.
Once a caller receives a user’s topic, on each subsequent visit within a week on the same web page (where the caller’s API is integrated) by the same user, only that specific topic is provided again.
Only if the user does not visit this web page for a week, a new random allocation of the known ("observed") topics assigned to this caller takes place.
If the user also visits other websites, they may receive additional topics (provided they have been classified as known).
Once a sufficient number of topics have been identified, the caller can receive up to three topics from the user, with one topic randomly selected from the top five topics each week.
Chrome allows users to view their categorization and delete irrelevant topics.
In addition, sensitive data categories (such as gender) should not be processed.
It is important to distinguish the Google Topics technology from the subsequent technical process of delivering ads.
Currently, Google provides an architecture for auctioning ad spaces through real-time bidding.
It is likely that Google will continue to support this approach, particularly as it provides its own product at all levels of the auction process.
However, it is technically possible to have a different configuration.
As the Google Topics technology focuses on inferring user interests and does not address the subsequent process, it is consistent that the Google Topics documentation does not provide any information on this matter.
\\ \\
The biggest difference from FLoC is that Topics does not use any additional browsing behavior in addition to the website names to determine the topics.
This is done in order to make cross-site identification by fingerprinting more difficult or even impossible.
In addition, under FLoC, the options to opt-out and view personal topics were not available.
The question of whether tracking, as demonstrated with FLoC~\cite{ana-floc}, will be possible with Topics remains open for now, as the final number of topics is not yet known.
However, there is still the possibility of creating a sequence of different topics associated with a user over time, which could be used for user identification purposes.
The difference to cookies~\cite{bullshit} lies in a different architecture~\cite{peng-cookies,Art.29Datenschutzgruppe.2012,Schott.2014}.
The integration of cookies is not only open to the respective website operators but also to third parties, whereas for FLoC and especially Topics, the implementation of the algorithm for determining topics is centrally managed by the browser operator Google.
Furthermore, on the basis of the existing documentation, cross-site tracking and the use of additional information about the recipients of advertisements (such as the query of the person on a "Demand Management Platform"~\footnote{See here: Herbrich/Niekrenz, CRi 2021, 129 (131).}) are not possible in principle.

\section{Legal assessment of Topics}
Section 25 TTDSG and the GDPR are of particularly relevance for the legal assessment of Topics.~\footnote{In addition, questions of cartel and competition law are relevant, which are not dealt with in this paper.}
Insofar as Section 25 TTDSG is applicable, it takes precedence over the GDPR pursuant to Art. 95 GDPR~\footnote{On the state of dispute regarding the interpretation of Art. 95 GDPR, see Gierschmann, MMR 2023, 22 with further references, Karg, in: Simitis/Hornung/Spiecker gen. Döhmann, Datenschutzrecht, 2019, Art. 95 Rn. 15 et seq.}, as Section 25 TTDSG is a sector-specific provision that transposed Art. 5 (3) ePrivacy Directive~\footnote{Directive 2002/58/EC of the European Parliament and of the Council of 12 July 2002 concerning the processing of personal data and the protection of privacy in the electronic communications sector (Directive on privacy and electronic communications).} and serves to protect the protect the integrity of the terminal equipment\footnote{cf. Art. 5 ePrivacy Directive.}.
On the other hand, the GDPR is applicable when the accessing entity processes the information, which are also personal data, after it has been stored and accessed.

\subsection{Compatibility of Topics with Section 25 TTDSG}
Section 25 TTDSG needs to be applicable to Topics and if this is the case, either one of the exceptions of Section 25 (2) TTDSG applies or the addressee of the provision must obtain consent pursuant to Section 25 (1), sentences 1 and 2 TTDSG.

\subsubsection{Applicability of Section 25 TTDSG to Topics}
\hfill \\ \\
As of July 2023, the Topics API is available in Germany. By updating Chrome with the Topics API, it constitutes a service provision within the territorial scope of TTDSG according to Section 1 (3) TTDSG~\footnote{Piltz, CR 2021, 555 (557); Riechert, in: Riechert/Wilmer, TTDSG, 2022, § 1 marginal no. 40.}.
The technology-neutral phrasing of Section 25 TTDSG covers not only cookies but also other previously unknown technologies, provided that information are stored or accessed in terminal equipment~\footnote{Schumacher/Sydow/von Schönfeld, MMR 2021, 603 (604); Schwartmann/Benedikt/Reif, RDV 2020, 231 (234); on the technical background of online tracking technologies see Möller, VuR 2022, 449 (450).}.
As Section 25 TTDSG does not explicitly name the addressees, but pertains to anyone who performes the corresponding action~\footnote{Riechert, in: Riechert/Wilmer, TTDSG, 2022, § 25 marginal no. 10; Schneider, in: Assion, HK-TTDSG, 2022, § 25 marginal no. 17.}, it appliesbroadly and is not restricted to telemedia service providers~\footnote{BayLfD, Bayerische öffentliche Stellen und Telemedien, Erläuterungen zum neuen Telekommunikation-Telemedien-Datenschutz-Gesetz (TTDSG), Orientierungshilfe, 2021, para. 24.}.
\\ \\
This section is applicable if information are stored in the terminal equipment acording to Section 2 (2) No. 6 TTDSG~\footnote{Nebel, CR 2021, 666 (668); Ettig, in: Taeger/Gabel, GDPR - BDSG - TTDSG, 4th ed. 2022, Section 25 TTDSG marginal no. 20; Riechert, in: Riechert/Wilmer, TTDSG, 2022, Section 25 marginal no. 12.} of an end user according to Section 2 (1) TTDSG in conjunction with Section 3 No. 13 TKG~\footnote{If the TTDSG does not provide for a definition of its own, it allows the use of standards of the TKG and TMG; Riechert, in: Riechert/Wilmer, TTDSG, 2022, § 25 marginal no. 14 considers the definition to be unsuitable for limiting or describing the scope of application of § 25 TTDSG; BT-Drs. 19/27441, p. 38; BayLfD, TTDSG Orientierungshilfe (Fn. 9), marginal no. 30; Nebel, CR 2021, 666 (669); for further problems in determining the end user in the case of several users see. Riechert, in: Riechert/Wilmer, TTDSG, 2022, § 25 marginal no. 13 ff.} or if information that are already stored there, is accessed~\footnote{Schumacher/Sydow/von Schönfeld, MMR 2021, 603 (604); Schwartmann/Benedikt/Reif, RDV 2020, 231 (234); for technical background see Möller, VuR 2022, 449 (450).}.
Neither the TTDSG nor the underlying ePrivacy Directive offers a definition of the terms “storage” or “access”.
However, with regards to “storage”, Section 3 (4) No. 1 German Federal Data Protection Act old version (BDSG a. F.), which transposed the Data Protection Directive (Directive 95/46/EC~\footnote{Directive 95/46/EC of the European Parliament and of the Council of 24 October 1995 on the protection of individuals with regard to the processing of personal data and on the free movement of such data.})~\footnote{Stoklas, ZD-Aktuell 2022, 00017.}, can be referred to.
According to this, "storage" refers to the collection, recording or preservation of personal data on a data carrier~\footnote{Roßnagel, in: Simitis/Hornung/Spiecker gen. Döhmann, Datenschutzrecht, 2019, Art. 4 Nr. 2 Rn. 19; crit.: Manaigo, K\&R 2022, 808 (811).}, which in this case, is the terminal equipment of the end user~\footnote{For a restrictive interpretation of the concept of storage based on the protective purpose of Section 25 TTDSG, see. Manaigo, K\&R 2022, 808 (812).}.
There is no clear definition of the term “access”.
However, based on the wording and the telos of Section 25 TTDSG, it is necesarry to perform an action ~\footnote{Manaigo, K\&R 2022, 808 (811).}, that results in the acquisition of stored information by breaching the integrity of the terminal equipment.
The identity of the person or party who stored the information in the terminal equipment bears no relevance~\footnote{Nebel, CR 2021, 666 (670).}.
If information is leaked to a third party without or against the data subject´s will, their privacy is compromised~\footnote{Cf. EC 24, 25 ePrivacy Directive.}.
If he end user itself voluntary disclosures their information (such as when accessing a website that transmits their IP address), there is no violation of their will and therefore no access in the sense of Section 25 (1) sentence 1 TTDSG~\footnote{Manaigo, K\&R 2022, 808 (811).}.
\\ \\
By inferring the topics through the Chrome browser using the implemented algorithm and storing them, Google is “storing” information on terminal equipment in accordance with Section 25 (1) sentence 1 alt. 1 TTDSG. In this regard, the algorithm performing the storage process is inconsequential for legal evaluation, as the software exclusively executes Google's code.  Additionally, the ability to infer topics solely through the installation and use of Chrome by end users is also irrelevant.Such actions cannot result in the denial of protection under Section 25 TTDSG, since Google's degree of action outweighs that of its end users.This aligns with Section 25 TTDSG's protective objectives, which safeguard the integrity of terminal equipment from unauthorised access. Failure to comply can jeopardise privacy, necessitating stricter measures under Section 25 TTDSG~\footnote{In contrast to Art. 6 of the GDPR, Section 25 of the TTDSG only provides for two exceptions to the requirement of consent, and no balancing clause comparable to Art. 6 (1) (1) (f) GDPR.}. 
Such non-compliance forms a foundation for advertising tracking and profiling. Browser providers do not appear to fall under Section 25 TTDSG initially because the designer of Google Topics did not anticipate the technical design of the legislator of the ePrivacy Directive. At this stage, it becomes apparent how cookies, especially third-party cookies, differ from each other. The user base is not restricted to a single actor. Rather, advertisers can employ the cookie technology or leverage data from other parties to display more pointed adverts. Consequently, there may be various recipients within the purview of Section 25 TTDSG~\footnote{The fact that in most cases the data is also transferred to Google because of the additional use of Google cookies is on another level; the problem cannot be discussed here.}.
\\ \\
The Caller accesses the topics within the meaning of Section 25 (1) sentence 1 alt. 2 TTDSG.
According to the current documentation on GitHub~\cite{githubTopics}, the Caller needs to request the topics so that they could be made available~\footnote{The following command is used for this:\\~\textbf{const Topics = await document.browsingTopics()}.}.
If the requirements for a "release" are met, the browser will make the topics available. This process occurs solely within the end user's browser.  Given this, one could argue that requesting the release does not violate end-user integrity since another action, specifically the release by the end user's browser, is necessary for any violation to occur. If the end user triggers the release by accessing the page, there is no action against their will, and therefore no access according to Section 25 (1) sentence 1 alt. 2 TTDSG.As a result, no legal basis is required for this access.However, the end user is constrained by a pre-defined programming that is beyond their control, necessitating the release of Topics, as direct browser access by third parties is not technically feasible.Given the minor prerequisites for releasing topics, topics will typically be made available to the caller.
 Therefore, the access should be allocated to the Caller based on evaluation, even if it does not directly compromise the integrity of the end user's device. Evaluation is necessary as the current legal criteria for determining access cannot account for technical reality. From an information technology perspective, obtaining information involves the end user, although this action is not legally relevant due to programming. The primary basis for the legal obligations is essentially rooted in information technology.
Furthermore, Google can be held accountable for the overall responsibility owing to the architecture it devised. This notion is not incompatible with the legislative intent, as the regulation was not formulated with such a scenario in mind.

\subsubsection{Legitimacy of storing and accessing Topics}
\hfill \\ \\
As a result, both Google, with respect to storage, and the callers, with respect to access, require a legal basis pursuant to Section 25 TTDSG. Section 25 (1) TTDSG contains the principle of consent and Section 25 (2) TTDSG contains two exceptions, neither of which is relevant to Topics. Section 25 (2) no. 1 TTDSG does not apply because the sole purpose of Topics is not the transmission of a message via a public telecommunications network. Section 25 (2) no. 2 TTDSG is also inapplicable because user-specific advertising and browser usage are to be regarded as two separate functions and the express request must also cover user-specific advertising, which is difficult to determine and, moreover, will rarely be the case in practice. Therefore, consent must be obtained from both Google and the Caller.
\\ \\
This consent pursuant to Section 25 (1) TTDSG in conjunction with Art. 4 (1) no. 11 and Art. 7 and 8 GDPR must be given voluntarily, for a specific case and in an informed and unambiguous manner.
In addition, the consent must take the form of a declaration or other unambiguous affirmative action by which the data subject indicates that he or she consents to the processing of their personal data~\footnote{On the prerequisites, see Klement, in: Simitis/Hornung/Spiecker gen. Döhmann, Datenschutzrecht, 2019, Art. 4 Nr. 11 Rn. 1 ff.}.
The ECJ stated in Planet49 that consent requires an action~\footnote{ECJ, C-673/17, ZD 2019, 556 m. Hanloser - Planet49.}, which Google has implemented in the final version of the Topics API. 
\\ \\
Consent must also be given in an informed manner. It is disputed whether it is necessary to name specific recipients or whether it is sufficient to name categories of recipients. This question was not decided by the ECJ in the "Planet49" decision, which only required the indication of "whether" third parties access the cookies, without dealing with the distinction between recipients and categories of recipients (Art. 10 (c) Directive 95/46/EC, Art. 13 (1) (e) GDPR) ~\footnote{ECJ, C-673/17, ZD 2019, 556, para. 75, 77 m. Hanloser - Planet49; different in. ECJ, judgment of 27.10.2022 - C-129/21, whereby this case refers to the more specific Art. 12 (1) ePrivacy Directive, which is why transferability to advertising purposes cannot be assumed, see: Schreiber/Dreesen, RDi 2023, 44.}.
The referring German Federal Court (Bundesgerichtshof, BGH) also did not determine the scope of the necessary information on the use of cookies~\footnote{BGH, I ZR 7/16, NJW 2020, 2540 - Cookie consent II (Planet49), para. 66.}.
However, as it is crucial for end users to know the recipients of the topics in order to be able to assess the risks of the data processing~\footnote{See Ettig in: Taeger/Gabel, GDPR - BDSG - TTDSG, 4th ed. 2022, § 25 TTDSG marginal no. 35.}, a specific designation of recipients is necessary~\footnote{See also EDSA, Guidelines 05/2020, para. 65; DSK, OH Telemedien, 2022, p. 36; Ettig in: Taeger/Gabel, GDPR - BDSG - TTDSG, 4th ed. 2022, § 25 TTDSG marginal no. 35 with further references.}.
This argument was used by the ECJ in case C-154/21 to argue that the recipients must be specifically named when asserting the right to information under Art. 15 (1) (c) GDPR~\footnote{ECJ, C-154/21 (RW v. Österreichische Post AG), EuZW 2023, 226 m. Sandhu; NJW 2023, 973 m. Petri; Markert, RGi 2023, 197.}.
This argument leads to a general principle, that can be transferred to the information obligations under Art. 13 GDPR (as well as Art. 14 GDPR)~\footnote{On this state of dispute see: Paal/Hennemann, in: Paal/Pauly, DS-GVO BDSG, 3rd ed. 2021, Art. 13 marginal no. 18; rejecting: Füllsack/Kirschke-Biller CR 2023, 103, 106 f.; Markert, RGi 2023, 197, 198.}.
This is also confirmed by the fact that specifying a category such as 'online advertisers' would contain more or less the same information as the purpose, which has to be specified anyway. If the additional requirement to provide information about the receiving party is not to be left empty in the case of cookies and Google Topics, it cannot in any case be limited to such a general form. Therefore, all recipients participating in the technology and thus potential recipients should be listed in the consent.
This should be done in a multi-layered approach, as it is frequently done currently~\footnote{Critical of various design variants Möller, VuR 2022, 449 (454).}.
This information must be provided even though the subsequent processing is not covered by Section 25 TDDSG, as it is decisive for the informed decision of the end user and the storage and access is the sole purpose for the further use of the information from the outset. It must be ensured that the end user does not have to reject each individual pre-selected recipient separately~\footnote{Report of the work undertaken by the Cookie Banner Taskforce, of 17.01.2023, paras 9, 10.}, as he might be tempted to accept all purposes if this is possible with a single click. This would lead to involuntary consent~\footnote{On the design of consent banners, see: DSK, OH Telemedien 2022, p. 33 f; on the question of the design of opt-out options see also: LG München, judgement of 29.11.2022, ref.: 33 O 14776/19; see on the design of cookie banners in general: Report of the work undertaken by the Cookie Banner Taskforce, published on 18.01.2023; EDSA, Guidelines 3/2022 on Dark patterns in social media platform interfaces: How to recognise and avoid them.}.
\\ \\
The aforementioned requirements must be met to by Google for storage and by the Caller for access.
Google must provide information about the purpose of the topics (personalised advertising), the duration of their storage and the specific recipients prior to their storage.
Since Google is not technically involved in the storage of the topics in the terminal, but is obliged to do so by virtue of its control of the entire process pursuant to Section 25 of the TTDSG, this should amount to obtaining consent when installing the browser or a corresponding update.
It seems problematic that Google may not yet be able to predict which Callers will participate as recipients of the topics at the time consent is obtained.
This is also an issue with cookies, but it may become more of a problem with Google Topics, as consent could already be obtained when the browser software is installed or updated.
This increases the lack of transparency for the end user and could only be solved by updating the information and consent.
\\ \\
Who the other potential recipients of the topics are depend on the design of the subsequent auction process.
Insofar as the Caller acts as an intermediary and passes on the topics to the advertising partners, these would also have to be listed in a multi-layer approach.
If, as is more likely and would correspond to the current auction process, the Callers have already agreed in advance with various advertising partners on which interests and at what price advertising is to be placed, there would be no need to pass on the topics.
In this case, Google would not have to provide further addressees, but would still have to provide information about the subsequent advertising process.
\\ \\
The Caller has a separate obligation to obtain consent to access the topics.
If the topics are to be passed on to advertisers, they must be informed of this.
In terms of implementation, consent could always be obtained before the topic is "requested".
This would mean that end users would always have to give their consent when accessing a page in a banner.
To avoid this step, consent could be obtained by Google when Chrome is installed.
This is supported by the fact that Google has to inform users about further use and Callers as part of the consent for storage anyway.
If a combined solution is chosen, care must be taken to ensure that it is clear to the end user that multiple consents have been given and what the implications are for them.

\subsection{Compatibility of Topics with the GDPR}
The subsequent processing steps are no longer covered by Section 25 TTDSG~\footnote{Piltz in: Plath, GDPR/BDSG/TTDSG, 4th ed. 2023, § 25 marginal no. 13; Grages, CR 2021, 834, 835.}, so the GDPR applies to the extent that personal data are processed.
Subsequent processing therefore requires a separate legal basis, which can be found in the GDPR~\footnote{Frenzel, in: Paal/Pauly, Ds-GVO BDSG, 3rd ed. 2021, Art. 6 marginal no. 7.}.
For this purpose, consent could also be obtained for subsequence processing pursuant to Article 6 (1) (1) (a) GDPR could be obtained for subsequent processing.
Google may request this consent together with the consent pursuant to Section 25 (1) TTDSG.
It must be ensured that the end user is clearly informed that he is giving more than one consent.
In principle, subsequent use is conceivable on the basis of the more flexible legitimate interest under Article 6 (1) (1) (f) GDPR, although this requires a comprehensive assessment of the mutual interests.
This assessment cannot be conclusively made on the basis of Google´s existing documentation.
Based on the literature on profiling for advertising purposes, it is likely to be rejected~\footnote{Guidance of the supervisory authorities on the processing of personal data for direct marketing purposes under the General Data Protection Regulation (GDPR), as of February 2022, p. 5 f.}.
Therefore, consent under Article 6 (1) (1) (a) GDPR may be required for subsequent processing, which could already be obtained at the time of storage by Google.

\section{Conclusion}
Topics is Google's latest approach to personalised advertising that does not involve the use of cookies and aims to enhance the protection of end-users' privacy through several technical measures. However, the Topics architecture also involves the storage and access of information on the device.
\\ \\
Under Section 25 TTDSG, Google is initially only obliged to provide information about the storage process, as the callers carry out the access and further processing. However, consent to storage, akin to cookies, should encompass additional usage purposes and recipients, without signifying consent to further data processing. Merely naming the recipient categories is insufficient. The caller's independent obligation would require obtaining the end user's consent before each access. Instead, it would also be acceptable - and more straightforward for all parties concerned - for Google to seek the callers' consent to gain access (in accordance with Section 25 TTDSG) and to permit further processing by Google (in accordance with the GDPR), along with consent to store the topics. In this instance, it must be explicitly stated that three consents can be connected to the declaration to guarantee lawful consent. 
\\ \\
If Google were to package the requisite consents, end users would only need to provide consent during the installation of the Chrome browser, as long as the callers identified at that time remain unchanged. Otherwise, any additional callers or changes to the procedure would require the consent to be updated accordingly. This caveat simplifies the content and upholds users' ability to control their information. The constant need to provide consent for cookies each time a website is visited results in users frequently opting for the easiest option and agreeing to all cookies, for the sake of being able to access the website rapidly. Consequently, they may grant permission for more cookies than they actually desire without receiving any notification about future data processing procedures and parties receiving their information. Centralised consent would reduce the burden on Callers as well as improve usability for end users, facilitating their ability to take note of provided information. 
\\ \\
Topics bring Google a step closer to achieving its goal of privacy-compliant and personalized advertising. The topics stored in the browser - and this marks the initial difference from cookies - do not afford unique identification of the end device, as they are not as unique even with a combination of multiple topics. Additionally, unlike cookies that can be utilized in diverse resources such as HTML, CSS, JavaScript, images, icons, and the like, hence facilitating tracking and profiling across pages, topics only inform interest groups when the corresponding code is executed to retrieve it. Additionally, Topics represents an improvement over FLoC as its pre-defined topics are less granular and the approach to topic aggregation differs. It is unclear whether additional information on individual user preferences can still be obtained.
\\ \\
The potential impact of the proposed procedure on the advertising market remains uncertain at this stage, and will depend Depending on the final technical design and the distribution or accumulation of individual roles, advertisers may become reliant on themes for user-specific advertising placement in the future. Any legal concerns (e.g. competition and antitrust law) should be discussed separately.

\bibliographystyle{splncs04}
\bibliography{main}

\end{document}